\documentclass[twocolumn,prl]{revtex4-1}\usepackage{graphicx,bm,times} \usepackage{natbib} \usepackage{miller,upgreek}
\begin{document}

\title{Epitaxial UN and $\upalpha$-U$_2$N$_3$ Thin Films}
\author{E. Lawrence Bright$^1$, S. Rennie$^1$, M. Cattelan$^2$, N.A. Fox$^{1,2}$, D.T. Goddard$^3$, and R. Springell$^1$}
\affiliation{$^1$School of Physics, University of Bristol, Tyndall Avenue, Bristol BS8 1TL. \\
$^2$School of Chemistry, University of Bristol, Cantocks Close, Bristol BS8 1TS. \\
$^3$National Nuclear Laboratory, Preston Laboratory, Springfields, Preston, Lancashire PR4 0XJ, UK.}

\begin{abstract}
Single crystal epitaxial thin films of UN and U$_2$N$_3$ have been grown for the first time by reactive DC magnetron sputtering.
These films provide ideal samples for fundamental research into the potential accident tolerant fuel, UN, and U$_2$N$_3$, its intermediate oxidation product.
Films were characterised using x-ray diffraction (XRD) and x-ray photoelectron spectroscopy (XPS), with XRD analysis showing both thin films to be [001] oriented and composed of a single domain.
The specular lattice parameters of the UN and U$_2$N$_3$ films were found to be 4.895\,\AA{} and 10.72\,\AA{}, respectively, with the UN film having a miscut of 2.6\,$^\circ$.
XPS showed significant differences in the N-1s peak between the two films, with area analysis showing both films to be stoichiometric.
\end{abstract}

\date{\today}

\maketitle

\section{Introduction}

Uranium mononitride, UN, is of significant interest to the nuclear industry due to its high melting point, high uranium density, and improved thermal conductivity in comparison to uranium dioxide, UO$_2$ \cite{Kurosaki2000}.
In addition to the enhanced thermal conductivity, providing an improved accident response, the 40\,\% higher uranium density of UN allows for lower enrichment or higher fuel burn-up \cite{Terrani2014}.
Despite these known advantages, there are still many material properties of UN yet to be fully investigated, in particular, there are concerns over the rapid oxidation of UN in water \cite{Dell1967,Sugihara1969,Jolkkonen2017}.
This oxidation reaction has been shown to progress with the formation of a U$_2$N$_3$ interlayer between UN and UO$_2$, making it also of interest \cite{Paljevic1975,Rao1991}.
A better understanding of this oxidation process, as well as fundamental fuel properties, such as thermal conductivity and irradiation performance, is required for UN to be considered as a viable accident tolerant fuel (ATF). 
Consequently, this area of research has had a renewal of interest, with several recent experiments utilising thin film samples \cite{Long2016,Wang2016,Lu2016}. 

Thin films provide an ideal way to research these properties, with their enhanced surface sensitivity being optimal for investigating surface reactions such as oxidation and hydrolysis, and ability to produce highly controlled samples, allowing for single variable investigations.
These experiments improve fundamental understanding of materials and provide experimental data comparable to theoretical calculations which are of particular importance in an area of research that is restricted as a result of radioactivity. 
In addition, thin films contain such little radioactive material that they do not require dedicated facilities and are more likely to be classed as exempt from radioactive material transport regulations.

Polycrystalline UN and U$_2$N$_3$ films have previously been grown by reactive DC magnetron sputtering and epitaxial thin films of UN$_2$ have been grown by polymer assisted deposition \cite{Black2001, Long2016,Scott2014}.
However, prior to this study there have been no reports on the successful deposition of epitaxial UN and U$_2$N$_3$ films. 
It is noted that while the fabrication of bulk single crystal UN is documented, there have been no prior reports of single crystal U$_2$N$_3$ \cite{Curry1965}. 
The ability to grow epitaxial UN and U$_2$N$_3$ thin films will therefore contribute to the advancement of ATF research, providing idealised samples on which to conduct fundamental material behaviour studies. 

\section{Experimental Details}

The films were grown in a DC magnetron sputtering system at the University of Bristol with 10$^{-8}$\,mbar base pressure, in-situ reflection high-energy electron-diffraction (RHEED), and substrate heating to 1200\,$^{\circ}$C, with the temperature at the substrate position calibrated using a pyrometer.
The system uses 5.5N argon at 7x10$^{-3}$\,mbar as the main sputtering gas, and houses a target of depleted uranium, producing deposition rates in the range of  0.5-1.5\,\AA/s.

A partial pressure of 5.5N N$_2$ is used to reactively deposit nitride films, with the pressure determining the phase deposited, as shown by Black $et$ $al.$ \cite{Black2001}.
Polycrystalline samples were grown at room temperature to optimise the N$_2$ partial pressure required to deposit single phase films of UN and U$_2$N$_3$, 2.0x10$^{-5}$\,mbar and 9.0x10$^{-4}$\,mbar, respectively, similar to that of Black $et$ $al.$

In order to grow single crystal films, compatible substrates with epitaxial matches were chosen and heated during deposition.
Substrates that did not contain oxygen were sought to prevent oxidation of the deposited nitride.
The substrates used were 10\,mm x 10\, mm, supplied by MTI corporation, single sided polished to 1-3\,\AA{} root mean square (RMS) roughness and mechanically mounted onto sample holders.

Cubic \hkl[001] CaF$_2$ was used as the substrate to epitaxially deposit U$_2$N$_3$ in the \hkl[001] direction at 700\,$^{\circ}$C.
It was selected as its bulk lattice parameter of 5.463\,\AA{} has only a 2.3\,\% mismatch with bulk $\upalpha$-U$_2$N$_3$, with bulk lattice parameter of 10.678\,\AA{} \cite{Swanson1953, Rundle1948}.

Bulk UN is cubic with a lattice parameter of 4.890\,\AA{} and was matched to Nb in the \hkl(001) plane with a 1:$\sqrt{2}$ relation and 45\,$^{\circ}$ rotation, Nb also being cubic with a lattice parameter of 3.300\,\AA{} \cite{Williams1959,Barns1968}.
UN \hkl[001] was grown on a Nb \hkl[001] buffer layer on a Al$_2$O$_3$ \hkl[1-102] substrate, with the Nb layer acting as both a chemical buffer, protecting the UN layer from oxidation, and physical buffer, improving the epitaxial match.
The Nb buffer and UN film were deposited at 800\,$^{\circ}$C and 500\,$^{\circ}$C, respectively.

All samples were capped with a 5\,nm layer of polycrystalline Nb or Au, deposited at room temperature, to prevent oxidation of the uranium nitride layers.

X-ray diffraction (XRD) and x-ray reflectivity (XRR) measurements were performed using a Philips X'Pert diffractometer with a Cu-K$\upalpha$ source.
Specular and off-specular 2$\uptheta$-$\upomega$, $\upomega$ (rocking curves), and $\upphi$ (azimuthal rotation) XRD scans were performed to investigate the crystallinity and epitaxy of the deposited films. 
XRR was used to measure the thickness and roughness of film layers and determine deposition rates. 

XRD scans were fitted analytically using GenX software, which uses a differential evolution algorithm to optimise the fit \cite{Bjorck2007}.
The GenX reflectivity package, which models scattering length density as a function of depth, was used to fit XRR measurements and obtain layer thickness and roughness values.

X-ray photoelectron spectroscopy (XPS) measurements were performed at the Bristol NanoESCA facility, which employs a monochromatic Al x-ray source (1486.7\,eV) and a ScientaOmicron XPS Argus analyser, and has an overall energy resolution of less than 300\,meV using a pass energy (PE) of 6\,eV. 
The instrument houses a 0.5-1\,keV Ar sputter gun, which was used to remove the capping layer on samples before taking measurements.
Survey scans were taken with a PE of 50\,eV, before scans of the N-1s and U-4f states were taken with a PE of 6\,eV.
Peaks were calibrated using the Fermi edge and further analysed using the CasaXPS software \cite{casaxps}.  

\section{Results}

\subsection{Structural Characterisation}

\begin{figure} \includegraphics[width=1\linewidth]{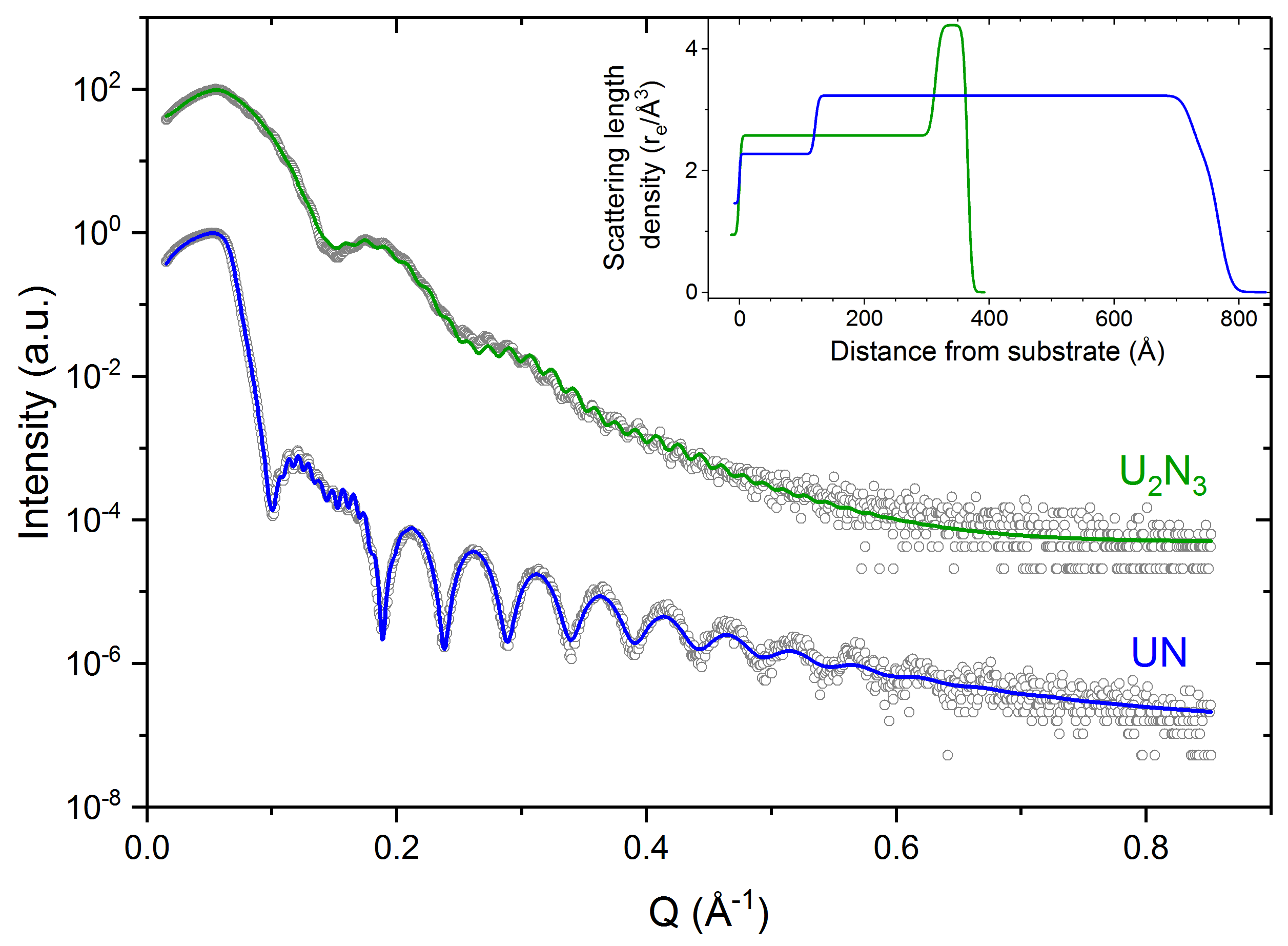} \caption{XRR scans and fits of the U$_2$N$_3$ and UN samples, shown in green and blue, respectively, with the scattering length density plot obtained from the fit inset. \label{fig:refs}} \end{figure}

The XRR measurements and fits of the \hkl[001] U$_2$N$_3$ and UN samples are shown in Figure \ref{fig:refs}.
XRR data was fitted by modeling electron density as a function of depth though the sample, as shown in the inset in Figure \ref{fig:refs}.
From the fits, it was found that the \hkl[001] U$_2$N$_3$ sample comprised of a 310\,\AA{} U$_2$N$_3$ layer and 50\,\AA{} Au cap, whereas the \hkl[001] UN sample was found to have a 600\,\AA{} UN layer, and 120\,\AA{} and 40\,\AA{} Nb buffer and cap respectively.
These values and the RMS roughnesses of each layer can be found in Table \ref{table}.

\begin{table*}[ht]
\caption{Sample characterisation results.}
\label{table}
\begin{tabular}{p{2.2cm}|p{2.2cm}p{2.2cm}p{2.2cm}p{2.2cm}p{2.2cm}p{2.2cm}}
\hline Sample &  Material & Layer &  Thickness (\AA) & Roughness (\AA) & Orientation & $\Updelta \upomega$ ($^{\circ}$) \\ \hline
U$_2$N$_3$ & CaF$_2$  & substrate  & -        & 2.7          & \hkl[001]                   & 0.12       \\
     & U$_2$N$_3$  & film       & 310      & 6.8          & \hkl[001]                   & 0.03       \\
     & Au    & cap           & 50       & 5.4          & polycrystalline                & -          \\ \hline
UN   & Al$_2$O$_3$ & substrate        & -        & 1.6         & \hkl[1-102]               & 0.04       \\
     & Nb    & buffer         & 120      & 4.3          & \hkl[001]                  & 1.22       \\
     & UN    & film    & 600      & 14.2          & \hkl[001]                   & 1.73       \\
     & Nb    & cap           & 40       & 15.0          & polycrystalline                & -        \\ \hline
\end{tabular}
\end{table*}

\begin{figure} \includegraphics[width=1\linewidth]{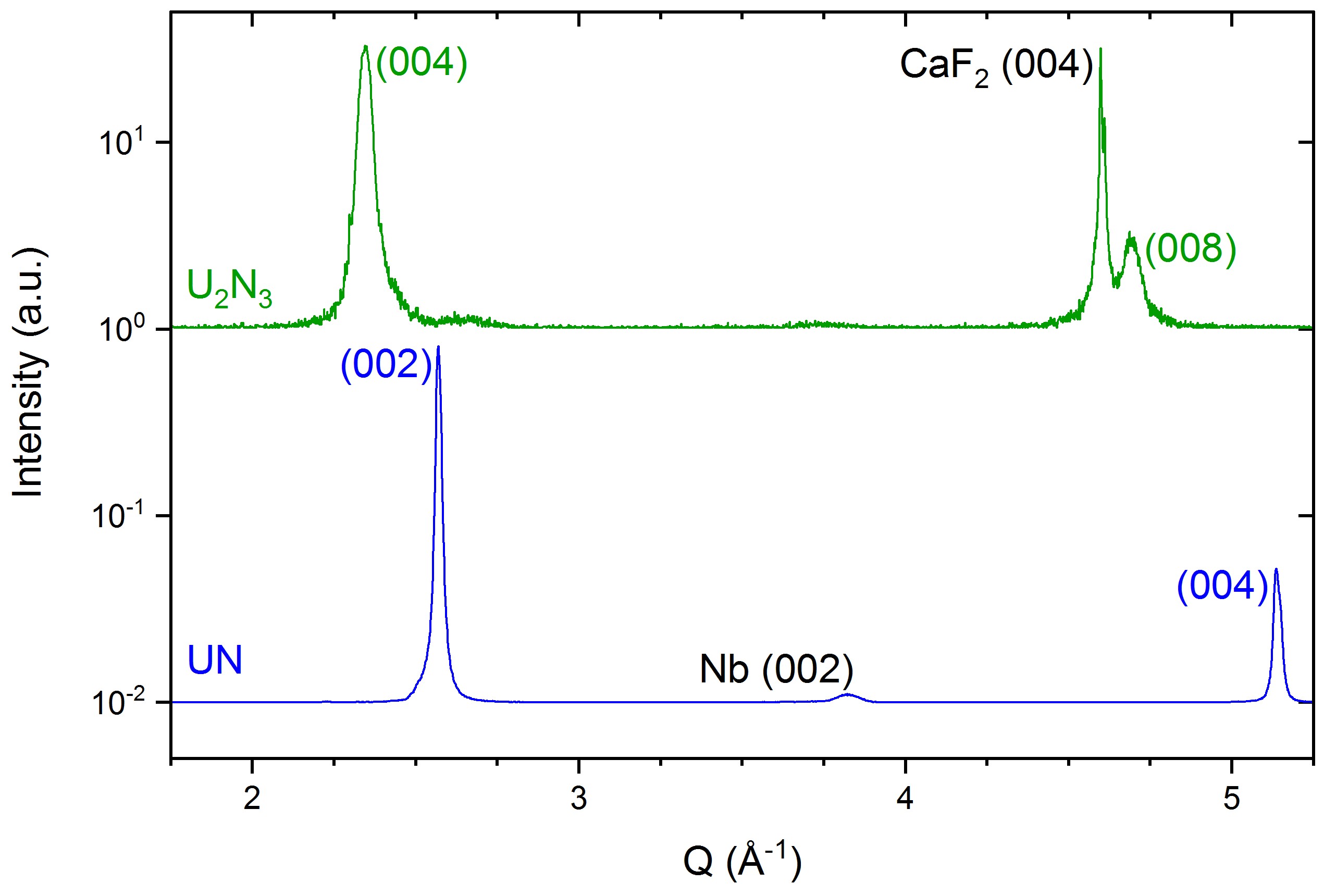} \caption{Specular 2$\uptheta$-$\upomega$ XRD scans of of the U$_2$N$_3$ and UN samples, shown in green and blue, respectively.\label{fig:ha}} \end{figure}

Figure \ref{fig:ha} shows the specular 2$\uptheta$-$\upomega$ XRD scans of the \hkl[001] U$_2$N$_3$ and UN samples, aligned to the specular film peaks.
It can be seen that in U$_2$N$_3$ film, grown on CaF$_2$, only the \hkl(004) and \hkl(008) reflections of U$_2$N$_3$ and \hkl(004) reflection of CaF$_2$ are visible, showing the film is highly oriented in this direction.
The same is true of the UN film grown on a Nb buffer on Al$_2$O$_3$, with only the \hkl(002) and \hkl(004) reflections of UN and \hkl(002) reflection of Nb visible.
From these reflections, it was calculated that the U$_2$N$_3$ $c$ lattice parameter (in the specular direction) is 10.72$\pm$0.01\,\AA{} and the UN $c$ lattice parameter is 4.895$\pm$0.001\,\AA{}.

\begin{figure} \includegraphics[width=1\linewidth]{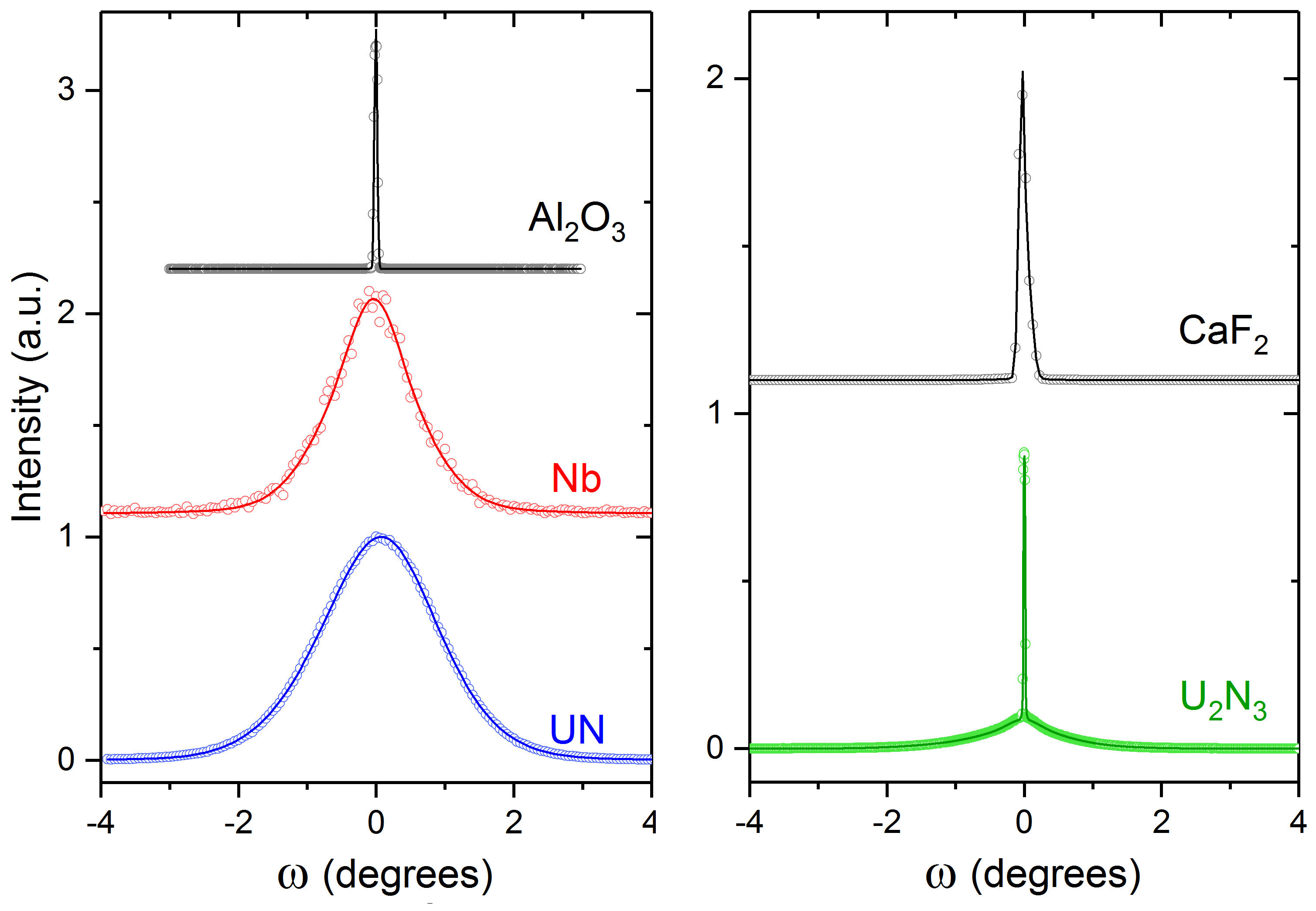} \caption{Rocking curves of the specular Al$_2$O$_3$ \hkl(024), Nb \hkl(002), and UN \hkl(002) Bragg peaks in the UN sample, shown in black, red, and blue, respectively, on the left. Rocking curves of the specular CaF$_2$ \hkl(004), and U$_2$N$_3$ \hkl(004) Bragg peaks in the U$_2$N$_3$ sample, shown in black and green, respectively, on the right.\label{fig:rc}} \end{figure}

The rocking curves or $\upomega$ scans of specular reflections in both the U$_2$N$_3$ and UN samples are shown in Figure \ref{fig:rc}, and the FWHM ($\Updelta \upomega$) of the fits shown in Table \ref{table}.
The rocking curve of the \hkl(004) U$_2$N$_3$ reflection is very sharp, with a FWHM of only 0.03\,$^{\circ}$, even narrower than the FWHM of the CaF$_2$ \hkl (004) substrate curve of 0.12\,$^{\circ}$.
A low intensity broad component is also present in the is curve, but not seen in that of the CaF$_2$ substrate, showing that there are areas of the U$_2$N$_3$ layer not completely commensurate with the substrate.

In the UN sample, both the Nb buffer and UN layer have broad rocking curves of 1.22\,$^{\circ}$ and 1.73\,$^{\circ}$ respectively, while the substrate curve is much narrower.
These large values show that the film layers are not in complete registry with the layer below.

\begin{figure} \includegraphics[width=1\linewidth]{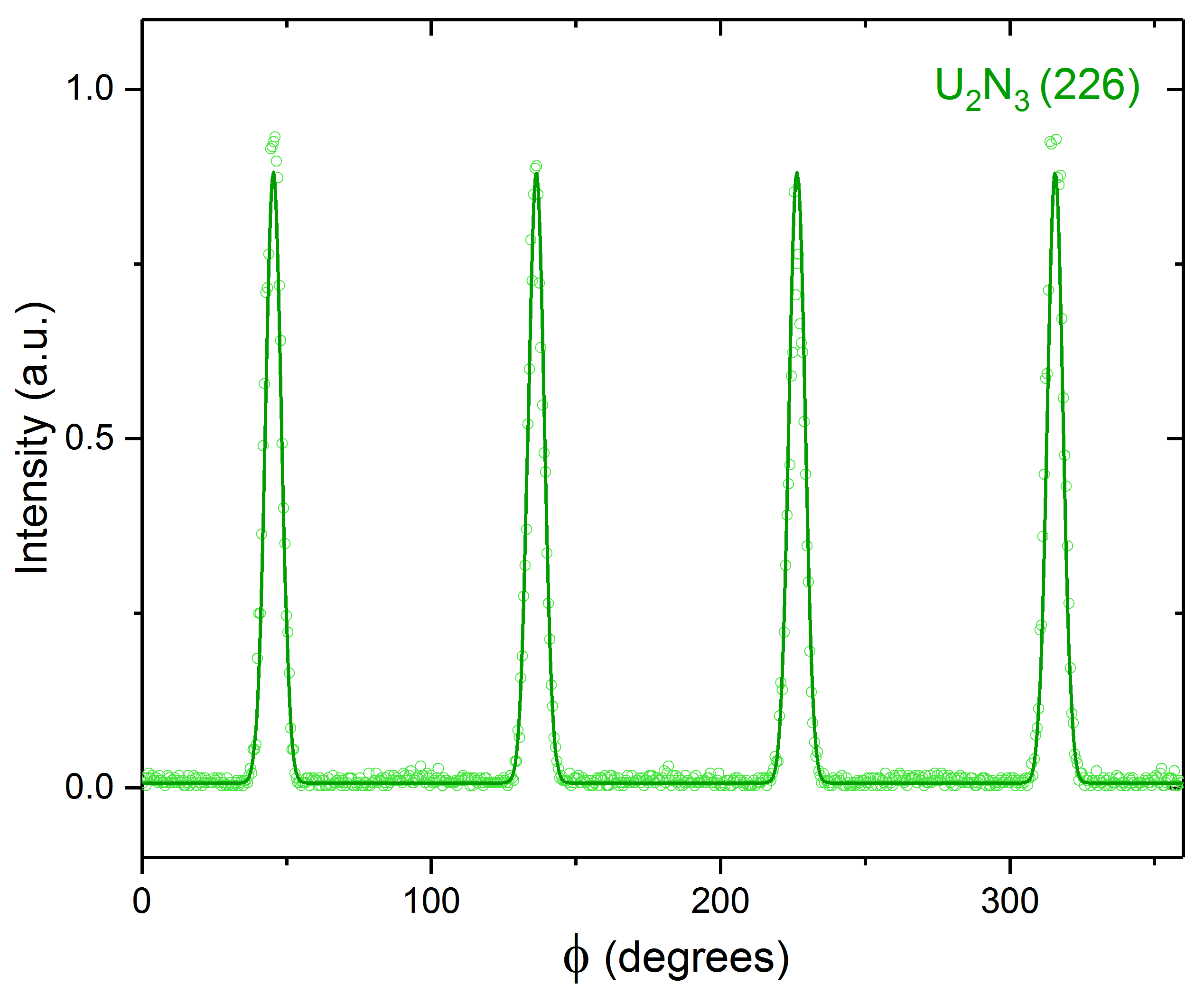} \caption{Phi scan of the off-specular U$_2$N$_3$ \hkl(226) Bragg peaks. \label{fig:u2n3phi}} \end{figure}

\begin{figure} \includegraphics[width=1\linewidth]{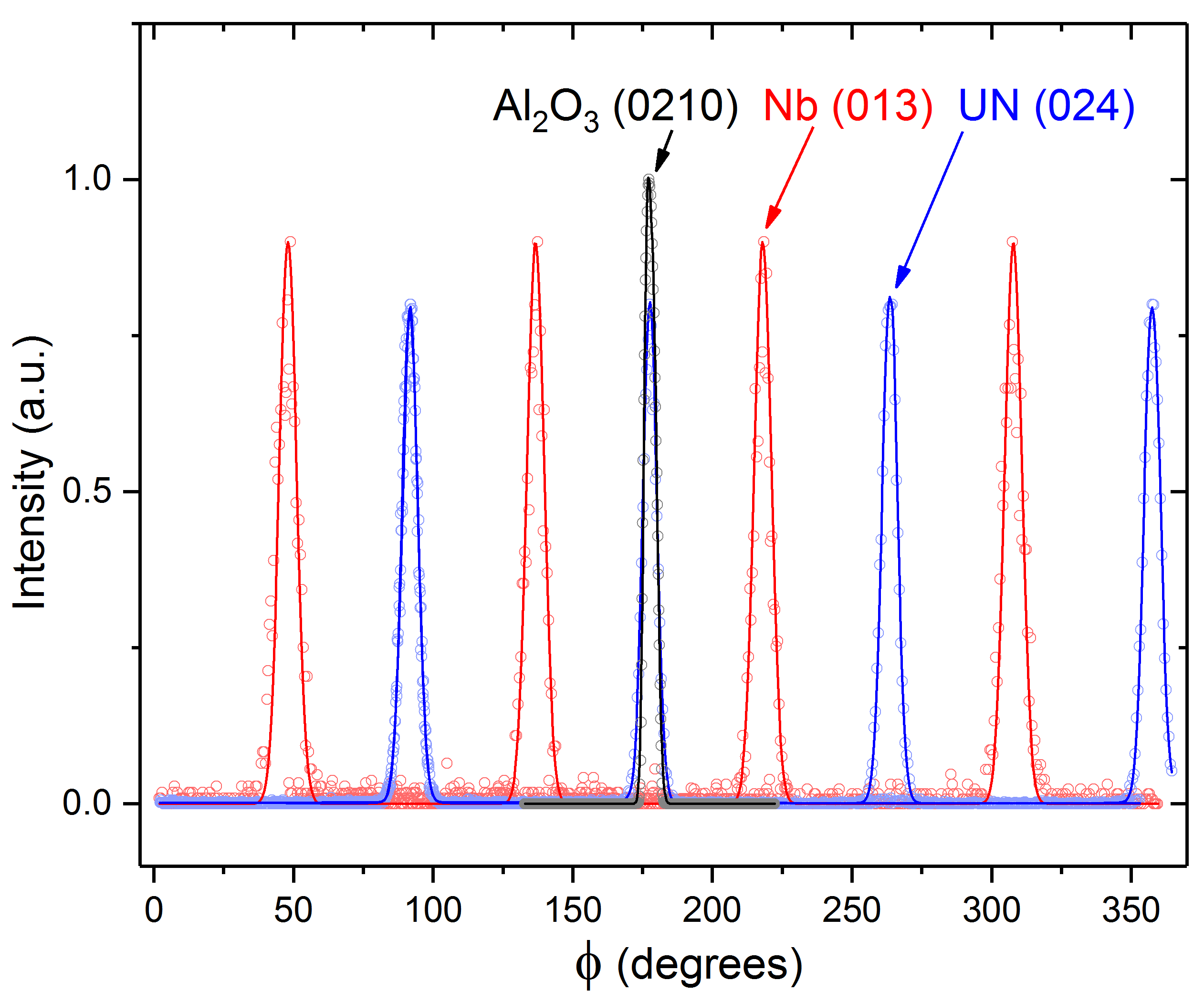} \caption{Phi scans of the off-specular Al$_2$O$_3$ \hkl(0 2 10), Nb \hkl(013), and UN \hkl(024) Bragg peaks, shown in black, red, and blue respectively. Due to the large miscut, each peak was scanned individually and normalised.\label{fig:unphi}} \end{figure}

While the specular XRD scans shows only the orientation of this film perpendicular to the surface plane, the in-plane orientation of the \hkl[001] U$_2$N$_3$ and \hkl[001] UN samples can be seen in the $\upphi$ scans shown in Figures \ref{fig:u2n3phi} and \ref{fig:unphi} respectively.

\begin{figure} \includegraphics[width=0.6\linewidth]{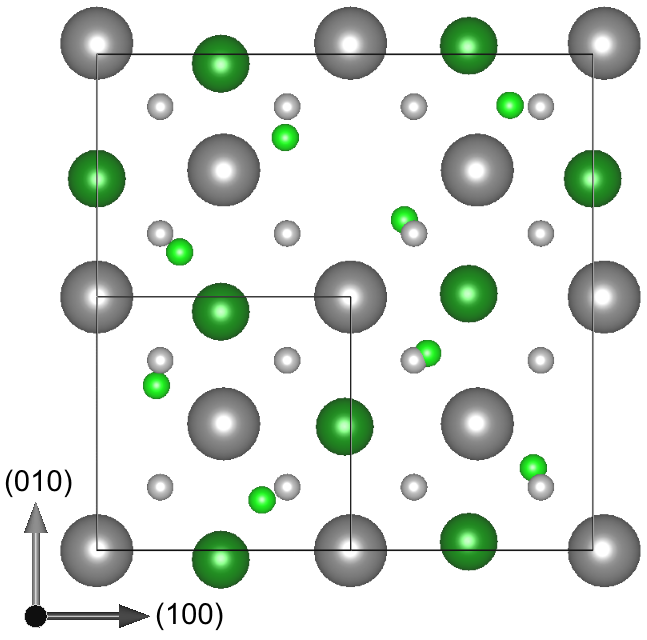} \caption{Model of \hkl(001) U$_2$N$_3$ on \hkl(001) CaF$_2$, with the uranium and nitrogen atoms shown as dark and light green and the calcium and fluorine atoms shown as dark and light gray. Made using VESTA software \cite{vesta}.\label{fig:u2n3match}} \end{figure}

The clear $\upphi$ dependence of the off-specular U$_2$N$_3$ \hkl(226) reflection shown in Figure \ref{fig:u2n3phi} indicates that there is a single domain present in the film.
Though not displayed, the off-specular CaF$_2$ reflections showed that the U$_2$N$_3$ film is oriented in the same direction as the substrate.
This is depicted in the model of the \hkl(001) planes of each of these in Figure \ref{fig:u2n3match}, which clearly demonstrates the 2:1 match between the two.

\begin{figure} \includegraphics[width=1\linewidth]{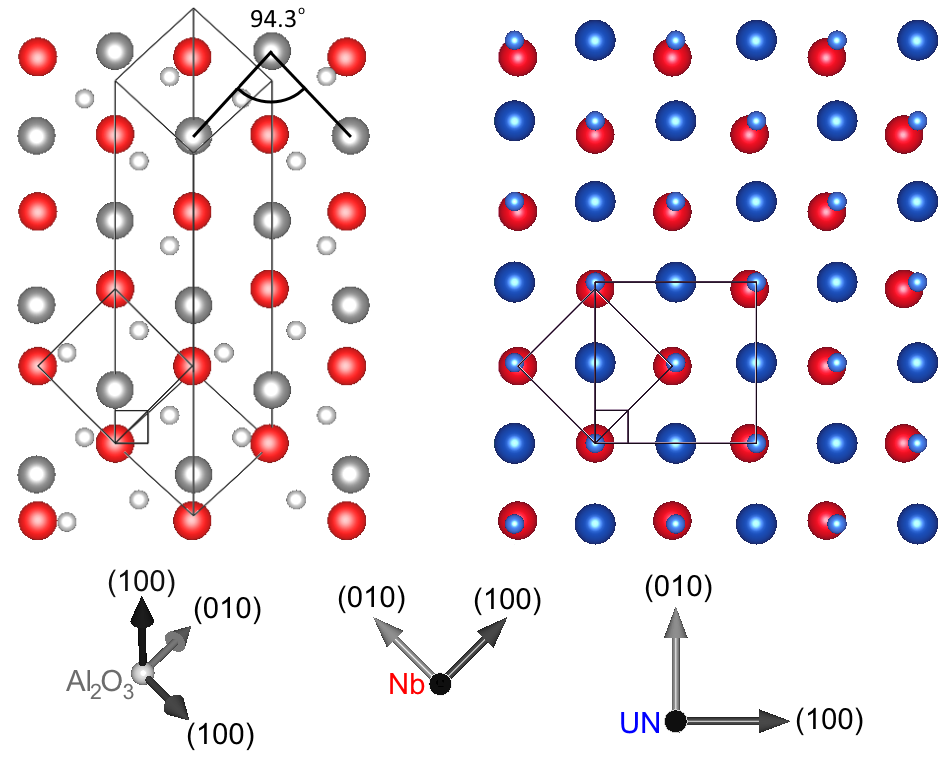} \caption{Model of \hkl(001) Nb on \hkl(1-102) Al$_2$O$_3$ and \hkl(001) UN on \hkl(001) Nb, with Al$_2$O$_3$ shown in gray, Nb in red and UN in blue.\label{fig:unmatch}} \end{figure}

Figure \ref{fig:unphi} shows that the \hkl(013) Nb and \hkl(024) UN Bragg peaks are dependent on the rotation of the sample, indicating that all crystallites in the \hkl[001] UN sample are of the same orientation.
Additionally this figure shows the orientational relationship between the Al$_2$O$_3$ substrate, \hkl[001] Nb buffer, and \hkl[001] UN film.
The 45\,$^{\circ}$ difference between the \hkl(013) Nb and \hkl(024) UN peaks indicates the $\sqrt{2}$ relationship between the buffer and film, as illustrated in the model in Figure \ref{fig:unmatch}.

This model shows the close match between the lattices of UN and Nb with a $\sqrt{2}$ relationship.
There also appears to be a close match between the Al$_2$O$_3$ lattice and Nb, however, the Al$_2$O$_3$ lattice in the \hkl[1-102] direction is not square, but rhomohedral, as can be seen by the 94.3\,$^{\circ}$ angle between 3 Al atoms shown in Figure \ref{fig:unmatch}.
As a rhombus can be considered a tilted square, it is likely this misfit is accommodated for by a tilt of the Nb crystal relative to the Al$_2$O$_3$ substrate such that the \hkl[001] Nb and \hkl[1-102] Al$_2$O$_3$ directions are not parallel.
This suggests there is a miscut between the Nb buffer and Al$_2$O$_3$ film, and as the UN film is matched to the Nb buffer, a miscut between the UN film and Al$_2$O$_3$ substrate.

\begin{figure} \includegraphics[width=1\linewidth]{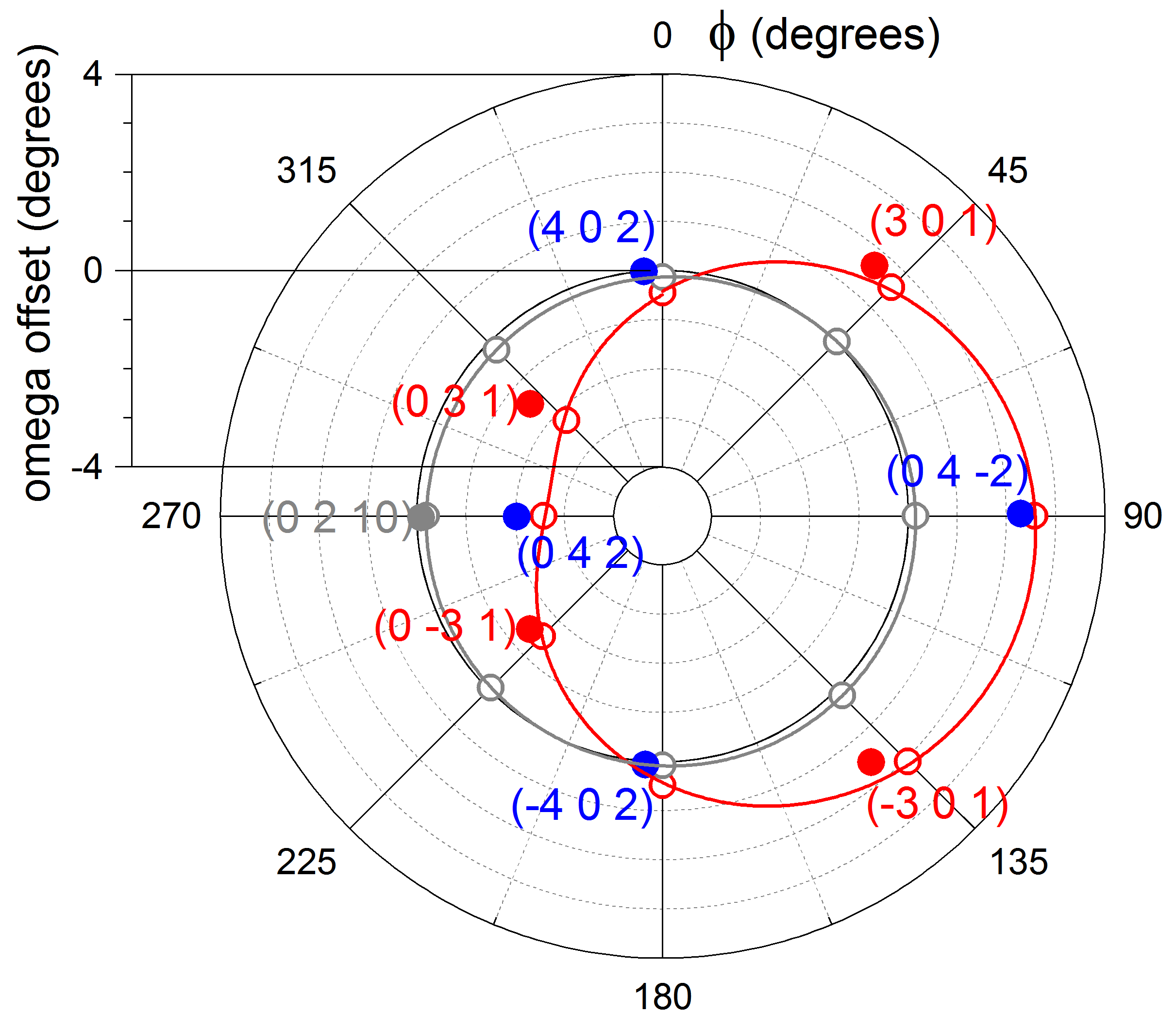} \caption{Omega offset of Bragg peaks as a function of $\upphi$ for the UN \hkl[001] sample, with Al$_2$O$_3$ shown in grey, Nb in red and UN in blue. The open and closed points show specular and labeled off-specular peaks respectively, and lines show sine function fits.\label{fig:unoff}} \end{figure}

In order to measure this miscut, the omega offset (angle relative to the specular direction) of various Bragg peaks was measured as a function of $\upphi$, sample rotation; the results can be seen in Figure \ref{fig:unoff}.
This figure shows labeled off-specular peaks of Al$_2$O$_3$, Nb, and UN as closed gray, red, and blue points respectively.
Open points show the specular \hkl(1-102) Al$_2$O$_3$ and \hkl(002) Nb Bragg peaks, fitted to sine functions.
The amplitude of this sine function is only 0.2\,$^{\circ}$ for the Al$_2$O$_3$ substrate, showing only a very small miscut between the \hkl[1-102] direction and the surface normal of the sample.
However, the amplitude of the sine fit to the Nb specular peaks is 2.6\,$^{\circ}$, with the Nb and UN off-specular peaks also lying close to this fit, showing that there is a large miscut in the \hkl[001] Nb and \hkl[001] UN layers.

\subsection{Chemical Characterisation}

\begin{figure} \includegraphics[width=1\linewidth]{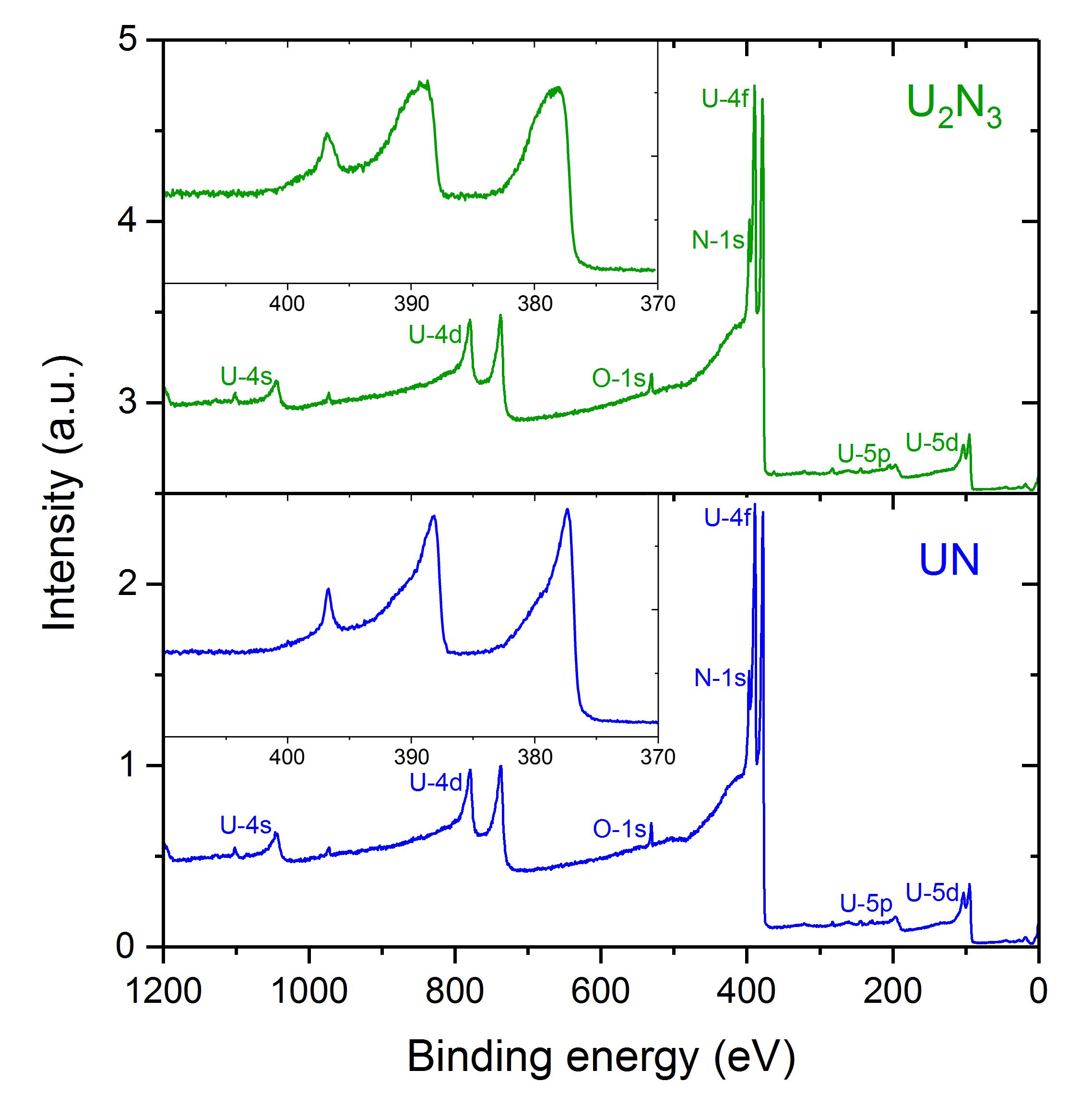} \caption{XPS survey scans of the \hkl[001] U$_2$N$_3$ and \hkl[001] UN samples, with U-4f and N-1s states inset.\label{fig:xpsall}} \end{figure}

Survey XPS scans taken after Ar sputtering of the \hkl[001] U$_2$N$_3$ and \hkl[001] UN samples, displayed in Figure \ref{fig:xpsall}, contain only peaks from U, N, and O contamination.
The lack of any peaks from the Nb and Au protective caps as well as Nb buffer and Ca and F substrate show that the spectra is being collected from the U$_2$N$_3$ and UN films only.
There are no visible C-1s peaks, showing the lack of carbon contamination in the films. 
However, the O-1s peak at 531\,eV is visible in both the U$_2$N$_3$ and UN films, showing oxygen contamination is present in both samples.

Spectra of the U-4f and N-1s states are inset in Figure \ref{fig:xpsall}, and show a clear asymmetry in the U-4f states.
This is more pronounced in the UN sample compared to the U$_2$N$_3$, and both the U-4f and N-1s peaks appear narrower in UN.

\begin{figure} \includegraphics[width=0.85\linewidth]{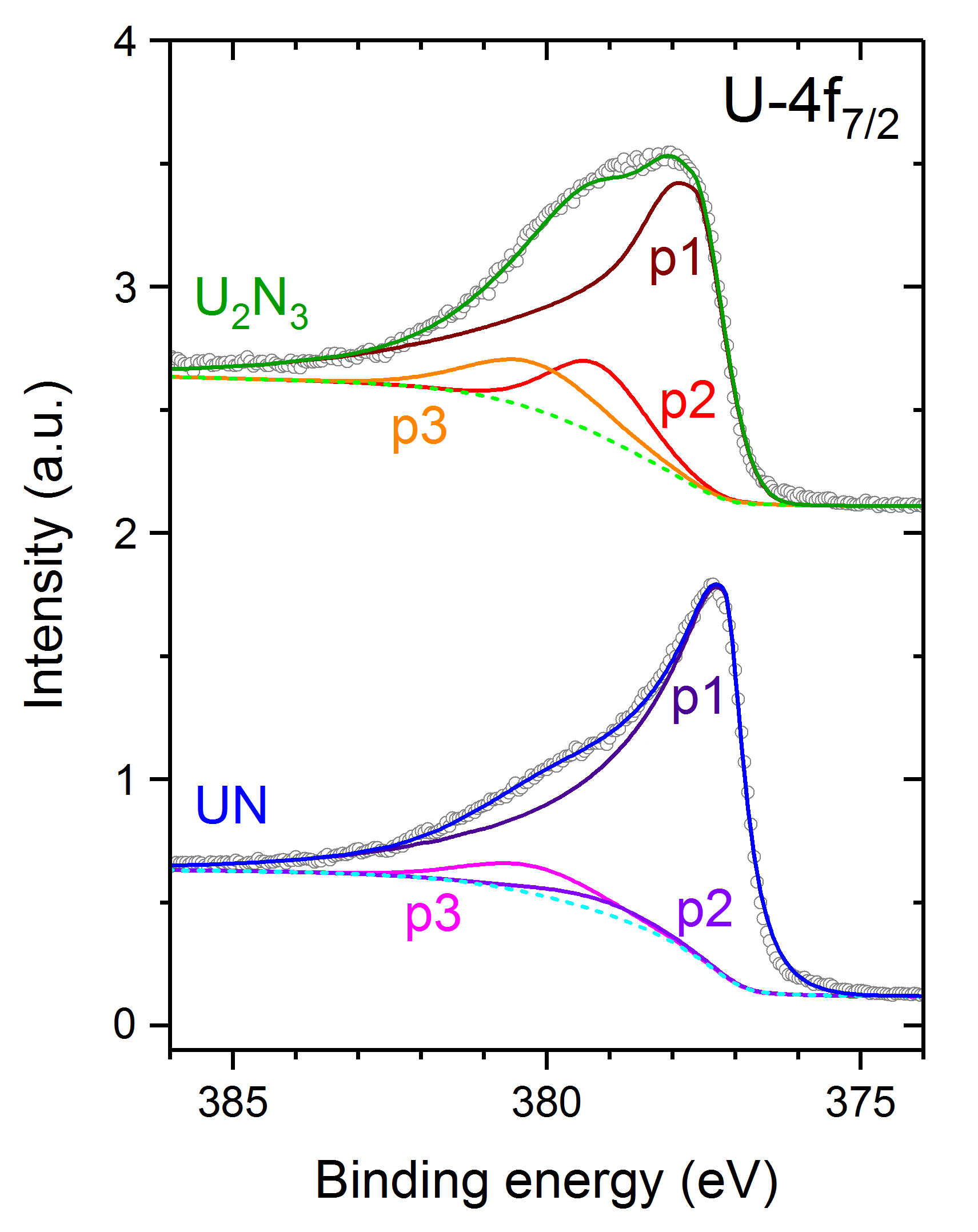} \caption{Fitted U-4f$_{7/2}$ spectra of the \hkl[001] U$_2$N$_3$ and \hkl[001] UN samples, with total fit shown in green and blue, respectively, and background shown by a dashed line. \label{fig:xpsufit}} \end{figure}

Fitting of the U-4f$_{7/2}$ peaks, plotted in Figure \ref{fig:xpsufit}, was performed using a Shirley background and Gaussian-Lorentzian product peaks, where \% GL is the percentage of Lorentzian weighting.
The lowest binding energy fitted peaks also contained an asymmetric exponential tail modifier, T, with this value and all others peak fit parameters displayed in Table \ref{xpstable}.

The fits showed the U-4f$_{7/2}$ state to be composed of two symmetric peaks at 379.2\,eV (p2) and 380.2-380.3\,eV (p3) and an asymmetric peak (p1) at lower binding energy for both U$_2$N$_3$ and UN.
For UN, this peak was fitted with a narrow FWHM of 0.8\,eV, higher asymmetry (low T), and 0.5\,eV lower binding energy compared to U$_2$N$_3$.
While the area and FWHM of p3 is similar in both UN and U$_2$N$_3$, p2 has a much more significant contribution to the U-4f$_{7/2}$ state in U$_2$N$_3$ than in UN.
The p3 peak, attributed to U(IV), along with the presence of an O-1s peak in the survey scan, show the presence of uranium oxide in the sample \cite{Allen1982}. 

\begin{figure} \includegraphics[width=0.85\linewidth]{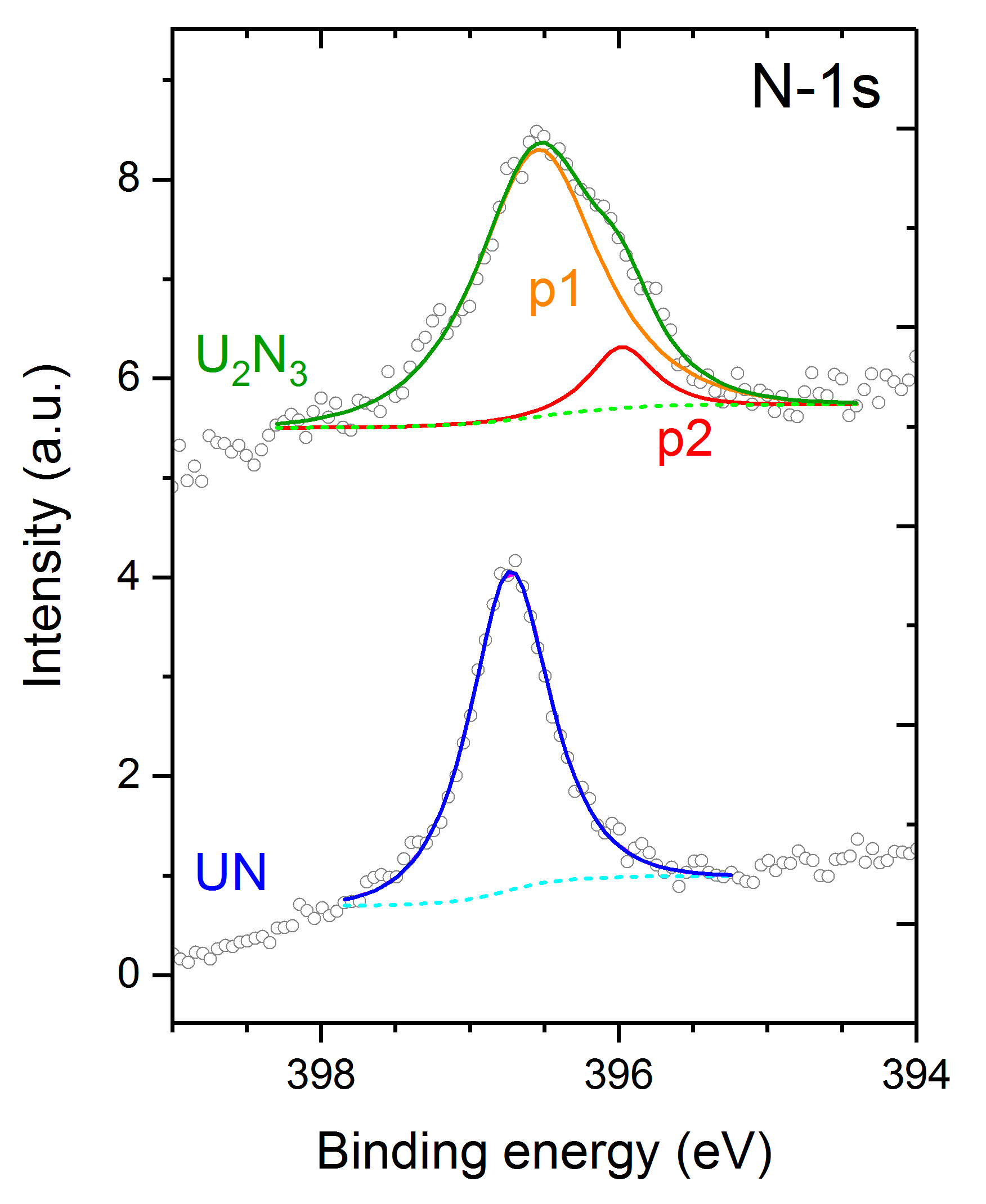} \caption{Fitted N-1s spectra of the \hkl[001] U$_2$N$_3$ and \hkl[001] UN samples, with total fit shown in green and blue, respectively, and background shown by a dashed line.\label{fig:xpsnfit}} \end{figure}

As with the U-4f$_{7/2}$ state, the N-1s states in U$_2$N$_3$ and UN were fitted with Shirley backgrounds and symmetric Gaussian-Lortenzian peaks, as shown in Figure \ref{fig:xpsnfit}.
Fitting of the N-1s state in U$_2$N$_3$ showed it to be composed of peaks at 396.6\, eV (p1) and 396.0\,eV (p2), with the former having a more significant contribution.
In the spectra from the UN sample, the N-1s state was fitted with only a single peak at 0.1\,eV higher energy than the main peak in U$_2$N$_3$.

The areas of the fitted peaks are shown in Table \ref{xpstable}, with the values normalised to the total area of the U-4f$_{7/2}$ peak for each sample.
Calculations of area ratios between the N-1s and U-4f (p1 and p2 only) were performed using cross sections given by Yeh $et$ $al.$, and gave values of 1.02$\pm$0.02 and 1.52$\pm$0.04 for UN and U$_2$N$_3$, respectively \cite{Yeh1985}.
The area of p3 in the U-4f$_{7/2}$ peaks was not included in the calculation as it is attributed to oxide in the sample.

\begin{table*}[ht]
\centering
\caption{X-ray photoemission line fit values.}
\label{xpstable}
\begin{tabular}{p{2cm}|p{1.4cm}p{1.4cm}p{2cm}p{2cm}p{2cm}p{2cm}p{2cm}}
\hline Sample &       Peak & & Position (eV) & FWHM (eV)  & \%GL & T & Area\\ \hline
U$_2$N$_3$   & U-4f$_{7/2}$ & p1   &    377.7     & 1.1       & 15     &  0.50 & 0.77 \\
       & U-4f$_{7/2}$ & p2   &     379.2     &  1.8        &  30    &  0 & 0.12 \\
       & U-4f$_{7/2}$ & p3   &  380.3        &  2.5        &  30    &  0 & 0.11 \\
       & N-1s & p1   & 396.6   & 0.9  & 80   & 0 & 0.07 \\
       & N-1s & p2   & 396.0   & 0.6   & 80   & 0 & 0.01\\ \hline
UN  & U-4f$_{7/2}$ & p1   & 377.2         & 0.8       & 90     & 0.38 & 0.90 \\ 
       & U-4f$_{7/2}$ & p2   & 379.2         & 2.3       & 30     & 0 & 0.03 \\
       & U-4f$_{7/2}$ & p3   & 380.2         & 2.5       &  30    &  0 & 0.07 \\
       & N-1s & p1   & 396.7   & 0.6 & 80   & 0 & 0.06 \\ \hline
\end{tabular}
\end{table*}

\section{Discussion}

The above results clearly show that epitaxial \hkl[001] UN and U$_2$N$_3$ single crystal thin films have been successfully grown for the first time.
XRD omega scans performed on the \hkl[001] U$_2$N$_3$ sample demonstrate that the film is in excellent registry with the CaF$_2$ substrate.
The $c$ lattice parameter of 10.72$\pm$0.01\,\AA{}, calculated from 2$\uptheta$-$\upomega$ scans, is higher than the bulk value of stoichiometric $\upalpha-$U$_2$N$_3$ of 10.68\,\AA{} \cite{Rundle1948}.
This difference could be caused by strain from the substrate and deviations from stoichiometry.

If this increase in the $c$ lattice parameter was caused by strain, a decrease in the $a$ and $b$ lattice parameter could be expected.
Unfortunately, the resolution limits of the x-ray diffractometer used and low number of accessible off-specular peaks do not allow for precise measurements of the in-plane lattice parameters and therefore strain in the film.

U$_2$N$_3$ is know to have a wide range of possible stoichiometries, with $x$ ranging from -0.2 to 0.5 in U$_2$N$_{3+x}$ \cite{Tagawa1974}.
While the literature on U$_2$N$_{3+x}$ is sparse, it is known that the lattice parameter of the cubic structure decreases with increasing value of $x$ \cite{Rundle1948}.
As the U$_2$N$_3$ film was deposited at the lowest pressure of N$_2$ found to deposit only single phase U$_2$N$_{3+x}$, and the $c$ lattice parameter is greater than that of stoichiometric U$_2$N$_3$, it is likely that $x$ is low. 
However, XPS area analysis suggests the sample is stoichiometric, within errors.

XPS of the U-4f valence states in the \hkl[001] U$_2$N$_3$ sample yielded results similar to those seen by Long $et$ $al.$, Wang $et$ $al.$, and Black $et$ $al.$ \cite{Long2016,Wang2016,Black2001}.
The U-4f$_{7/2}$ is peak found to be at 377.7\,eV, 0.5\,eV higher than in UN, and is fitted with an asymmetric peak at this energy, p1, and a symmetric peak at 379.2\,eV, p2, which is consistent with the analysis of Wang $et$ $al.$.
While the p1 peak is asymmetric, it is less so than that of UN and U metal, which along with the higher binding energy, show the partial localisation of the 5f states, as described by Black $et$ $al.$
The p2 peak could be attributed to U (III), as seen in U (III) oxyhalides\cite{Thibaut1982}.
The presence of oxygen contamination in the sample is shown by the small U (IV) peak at 380.2\,eV, p3, and the O-1s peak seen in the survey. 

While Wang $et$ $al.$ and Long $et$ $al.$ both claim that there is no difference in the N-1s state between UN and U$_2$N$_3$, the present data shows clear evidence of a shoulder at lower binding energy, as well as a 0.1\,eV shift in energy of the main component of the peak.
It is difficult to determine whether this is present in other literature, as most have lower resolution, and none fit the N-1s peak.
There is, however, literature on this second component in other metal nitride systems, such as TiN, where it appears under oxidation and is attributed to the formation of oxynitrides \cite{Glaser2007}. 
However, since the U (IV) peak (p3) in the U$_2$N$_3$ and UN samples are very similar, but this second N-1s component is not present in UN, it is likely that it is not due to oxidation but is instead an indication of the mixed states present in U$_2$N$_3$. 
The broader FWHM of the main component in the U$_2$N$_3$ N-1s peak of 0.9\,eV compared to the 0.6\,eV FWHM seen in UN are further evidence of the mixed states in U$_2$N$_3$.

XRD of the \hkl[001] UN sample showed it to be of a single domain, with a $c$ lattice parameter of 4.895$\pm$0.001\,\AA{}, close to bulk values, but of much lower quality than the \hkl[001] U$_2$N$_3$ sample.
This is evident in the broad rocking curves of both the UN film and Nb buffer layers, as well as the large miscut, which shows a lack of coherence between the Nb buffer and Al$_2$O$_3$ substrate.
As the UN film can only be as good quality as the Nb buffer, and Nb growth on \hkl[1-102] Al$_2$O$_3$ growth is shown to be optimised at 800\,$^\circ$C, it is unlikely that the quality of the UN film can be improved using this system \cite{Claassen1987}.
The miscut lying in same plane as the Al$_2$O$_3$ $c$ axis and specular direction is consistent with literature, which also shows the \hkl[111] Nb direction to align with the Al$_2$O$_3$ $c$ axis \cite{Gutekunst1997}.
While these papers also find the large miscut between Al$_2$O$_3$ and Nb, none provide the explanation of it arising from the accommodation of the rhombohedral Al lattice in the \hkl(1-102) plane.

Spectra of the U-4f states collected from the UN sample shows sharp asymmetric peaks at higher binding energy than U metal but lower than U$_2$N$_3$, which is comparable to the spectra of Norton $et$ $al.$, Long $et$ $al.$, Black $et$ $al.$, and Wang $et$ $al.$ \cite{Long2016,Wang2016,Black2001,Norton1980a}.
This is indicative of the itinerant nature of the system, as described by Fujimori  $et$ $al.$\cite{Fujimori2012}.
Slight differences in the spectra arise due to varying levels of oxide in each sample, seen by the U (IV) peak at 380.2\,eV.
Comparing to the only fitted spectra in the literature and the only spectra taken from a single crystal UN sample, that of Samsel-Czeka\l{}a $et$ $al.$, this work shows a much smaller contribution from the peaks at 379.2\,eV and 380.2\,eV, likely due to the higher purity of the present sample \cite{Samsel-Czekala2007}. 

\section{Conclusion}

Single crystal UN \hkl[001] and U$_2$N$_3$ \hkl[001] thin films have been successfully deposited via reactive DC magnetron sputtering.
XRD analysis shows that both the UN and U$_2$N$_3$ samples are single domain, with specular lattice parameters comparable to bulk values.
The U$_2$N$_3$ sample was shown to be of high quality, with good registry to the CaF$_2$ substrate, having a particularly narrow rocking curve.
The rocking curve of the UN sample was found to be significantly broader than its Al$_2$O$_3$ substrate, likely due to the large miscut between the substrate and buffer, however, off-specular measurements clearly demonstrate a single domain.
Chemical characterisation, conducted via XPS, show the presence oxygen contamination in the thin films.
The U-4f peaks were found to be highly asymmetric in the UN sample, indicative of its metallic nature.
This was observed to a lesser extent in the U$_2$N$_3$ film.
Additionally, the N-1s peak was found to differ between the UN and U$_2$N$_3$ samples, with the latter showing two broader components at lower binding energies.

\section{Acknowledgments}

The authors are grateful to G. H. Lander for helpful discussions.
The authors acknowledge access to the Bristol NanoESCA Facility under EPSRC Strategic Equipment Grant EP/M000605/1 and funding from EPSRC grant 1652612.

\section{References} \bibliography{cit}{} \end{document}